Title:

# Modified Stoney formula for obtainment of stress within thin films on large deformed isotropic circular plates

Haijun Liu [1,2,*], Minghui Dai [1], Xiaoqing Tian [1,2], Shan Chen [1,2], Fangfang Dong [1,2] and Lei Lu [3,*]

[1]School of Mechanical Engineering, Hefei University of Technology, Hefei 230009, P.R. China;

[2]Anhui Engineering Laboratory of Intelligent CNC Technology and Equipment, Hefei 230009, China

[3]Collaborative Innovation Center of Suzhou Nano Science and Technology, Soochow University, Suzhou 215021, P.R. China; jluc2ll@163.com

[*]Correspondence: liuhaijun@hfut.edu.cn (Haijun Liu); jluc2ll@163.com (Lei Lu)

**ABSTRACT**

Stoney formula is widely used to obtain the residual stress in the films on isotropic circular plates. However, significant errors would be introduced for large deformations because of the assumption of small deformation in the Stoney formula. In this study, a modified Stoney formula was proposed to extend its scope of application to the nonlinear domain. One-phase exponential decay function with coefficient $p$ was used to relate the curvature of the substrate to the stress in the film. The coefficient $p$ could be expressed as a function of the thickness, diameter, Young's modulus, and Poisson's ratio of the circular plate. The linear fitting technique was applied to ascertain the relationship between the coefficient $p$ and the aforementioned parameters. The simple yet accurate modified Stoney formula could be used to calculate the residual stress in the film directly from the measured curvature of isotropic circular plates with various dimensions and materials.

# 1. INTRODUCTION

For film/substrate systems, the Stoney formula has been widely used to calculate film stress.[1-4] The method is simple as the film stress could be obtained by basic arithmetic operations of the curvature of the deformed substrate without the need to know the material property of the films.[5] However, several assumptions are made in the derivation process of the Stoney formula, which limits the scope of its application. A considerable number of researches have been conducted to expand the application scope of the Stoney formula.

Janssen et al. proposed a modified form of the Stoney formula in which the elastic stiffness constants of monocrystalline silicon were used to overcome the assumption of isotropy of substrate materials.[4] Chen et al. presented a modified Stoney formula for film/substrate systems with different biaxial stresses in two axes on thin films.[6] Injeti et al. derived a modified Stoney formula for film/substrate systems with thin anisotropic substrates which did not meet the assumption that the substrate thickness was much larger than the film.[7] Qiang et al. proposed an extension of the Stoney formula for the incremental stress of the film which violated the assumption that the stress is uniformly distributed in the film.[8]

Another hypothesis in the Stoney formula is that the deformations are infinitesimally small compared with the substrate thickness. The increase of the error would become more and more obvious with the increase of the residual stress and the error may even exceed 50% of the true residual stress.[9] Therefore, the application scope of the Stoney formula is severely limited. To address this limitation, Masters et al. proposed a method to relate the stress and substrate curvature by minimizing the total strain energy of the film/substrate system.[10] A nonlinear analytical model was also proposed by Schicker et al. to describe the stress-curvature relationship.[11] One drawback of the energy method is that the solving of the fifth-order polynomial equation is needed which is often needed to be solved numerically. The finite element method is also adopted to simulate the deformation under specified residual stress.[9,12] The curvature could be easily obtained when the residual stress is assigned. However, the residual stress could not be directly obtained inversely from the curvature using the energy method or finite element models. The time-consuming trial and error method is often needed.

With the increase of residual stress in the film, the stress-curvature relationship would be nonlinear. We found that a one-phase exponential decay function with one coefficient was suitable to relate the curvature of the substrate to the stress in the film. In this study, we ascertained the relationship between the coefficient and the thickness, diameter, Young's modulus, and Poisson's ratio of the substrate. Film/substrate systems with different values of the parameters of the substrate dimensions and material properties were loaded with different residual stresses to determine the functional relationship between the coefficient and the substrate dimensions and material properties. Finally, a modified Stoney formula which was simple yet accurate was obtained to relate the curvature and the residual stress in the films.

## 2. OBTAINMENT METHODS OF STRESS-CURVATURE RELATIONSHIP

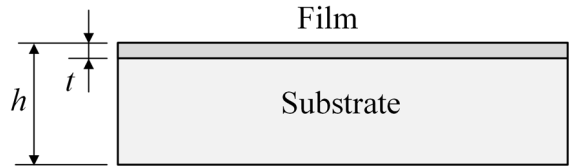

Fig. 1 Film/substrate system

### 2.1 Energy method

For a typical film/substrate system (Fig.1), the curvature and stress relationship could be obtained by minimizing the total strain energy of the system.[10] The substrate deflection could be expressed as

$$\omega(x, y) = \frac{1}{2}(a_0 x^2 + b_0 y^2) \qquad (1)$$

where $a_0$ and $b_0$ denote the curvatures of two perpendicular directions. It is a spherical shape when the value $a_0$ is equal to $b_0$. Otherwise, it is a cylindrical or saddle shape when the value $a_0$ is not the same as $b_0$.

The strain of the system could be expressed as

$$\varepsilon_x = a_1 + a_2 x^2 + a_3 y^2 - az, \quad \varepsilon_y = b_1 + b_2 y^2 + b_3 x^2 - bz, \quad \gamma_{xy} = (2a_3 + 2b_3 + a_0 b_0)xy \qquad (2)$$

where $a_1, a_2, a_3, b_1, b_2,$ and $b_3$ are unknown constants.

There are two layers in the film/substrate system. There exists initial residual stress $\sigma^*$ in the film and the strain energy density in the film could be expressed as

$$U_{density}^{film} = \frac{1}{2}Q_{11}\varepsilon_x^2 + Q_{12}\varepsilon_x\varepsilon_y + \frac{1}{2}Q_{22}\varepsilon_y^2 + \frac{1}{2}Q_{66}\gamma_{xy}^2 + \sigma^*\varepsilon_x + \sigma^*\varepsilon_y \qquad (3)$$

There does not exist residual stress in the substrate and the strain energy density in the substrate could be expressed as

$$U_{\text{density}}^{\text{substrate}} = \frac{1}{2}Q_{11}\varepsilon_x^2 + Q_{12}\varepsilon_x\varepsilon_y + \frac{1}{2}Q_{22}\varepsilon_y^2 + \frac{1}{2}Q_{66}\gamma_{xy}^2 \quad (4)$$

For isotropic materials, the reduced elastic stiffness constants are expressed as

$$Q_{11} = \frac{E}{1-v^2}, \quad Q_{12} = \frac{vE}{1-v^2}, \quad Q_{66} = \frac{E}{2(1+v)} \quad (5)$$

where $E$ and $v$ are the material's Young's modulus and Poisson's ratio, respectively.

The integration of the strain energy density of the film/substrate system along $z$-direction could be expressed as

$$U_{\text{density}}^z = \int_{-h/2}^{h/2-t} U_{\text{density}}^{\text{substrate}} \mathrm{d}z + \int_{h/2-t}^{h/2} U_{\text{density}}^{\text{film}} \mathrm{d}z \quad (6)$$

For circular plates, the integration of the strain energy density along $x$ and $y$ directions could be conducted in a cylindrical coordinate system by substituting $x$ with $r\cos\theta$ and $y$ with $r\sin\theta$.

The total strain energy density of the film/substrate could be obtained as:

$$U_{\text{total}} = \int_0^{2\pi} \int_0^R U_{\text{density}}^z r \mathrm{d}r \mathrm{d}\theta \quad (7)$$

To minimize the total strain energy, the derivative of $U_{\text{total}}$ with respect to all the unknown coefficients should be zero.

$$\frac{\partial U_{\text{total}}}{\partial a_i} = 0, \quad \frac{\partial U_{\text{total}}}{\partial b_i} = 0 \ (i = 0, 1, 2, 3) \quad (8)$$

There is only one real solution before the substrate bifurcates. The system presents as a spherical shape in this stage and $a_0$ is equal to $b_0$. When the system bifurcates, there are three real solutions and the middle of them corresponds to the spherical shape which is no longer stable. In this study, the curvatures were obtained using numerical calculation software (Maple 2018.0, Canada). For a film/substrate system, the stress and curvature relationship curve was shown in Fig. 2. Young's modulus of the system was 100 GPa and Poisson's ratio was 0.2. The thickness of the plate was 400 μm and the diameter was 0.2 m.

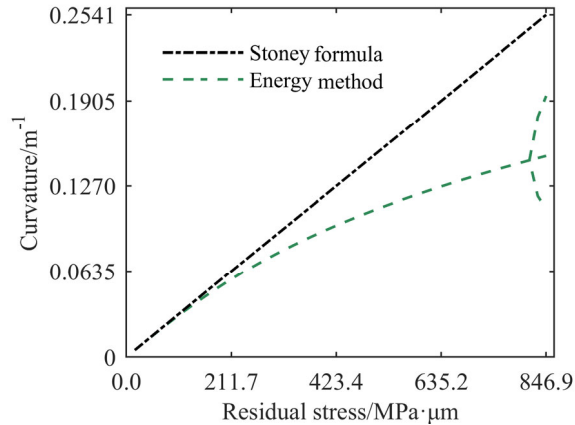

Fig. 2 Curvature-stress relationship using energy method

**2.2 Finite element method**

Finite element methods have turned to be a powerful tool to simulate the deformations of various thin plates with the development of computing power. Film/substrate systems are typical shell structures and shell elements in finite element software are suitable to predict the deformations under residual stress.[12,13] In this study, commercially available finite element software (Mechanical APDL 15.0, Ansys, USA) was adopted to relate the curvature and the residual stress in films. The element type SHELL181 was selected as it is suitable for analyzing thin shell structures and large deformation applications. The circular plate was meshed with quadrilateral elements and three nodes were selected to add displacement constraints to the plate which can deform freely but can't move or rotate as a whole.

The assumptions of the element type become fewer compared with the energy method. First, the transverse components of strained are included in the finite element method as the Mindlin-Reissner shell theory is adopted in the modeling of SHELL181 element rather than the classical Kirchhoff plate theory in the energy method. Second, the deformed shape of the film/substrate system was not simply represented by a fixed curvature. As shown in Fig. 3, the deformed plate was not a perfectly spherical shape due to the difference of stress states in different radial positions. There do not exist forces in the periphery of the plates in the finite element model while it is assumed the force conditions are all the same regardless of the radial positions in the energy method. Although the shape was not exactly a sphere it was still very close to spheres. Therefore, sphere fitting was conducted to the deformed data and a corresponding curvature could still be obtained.

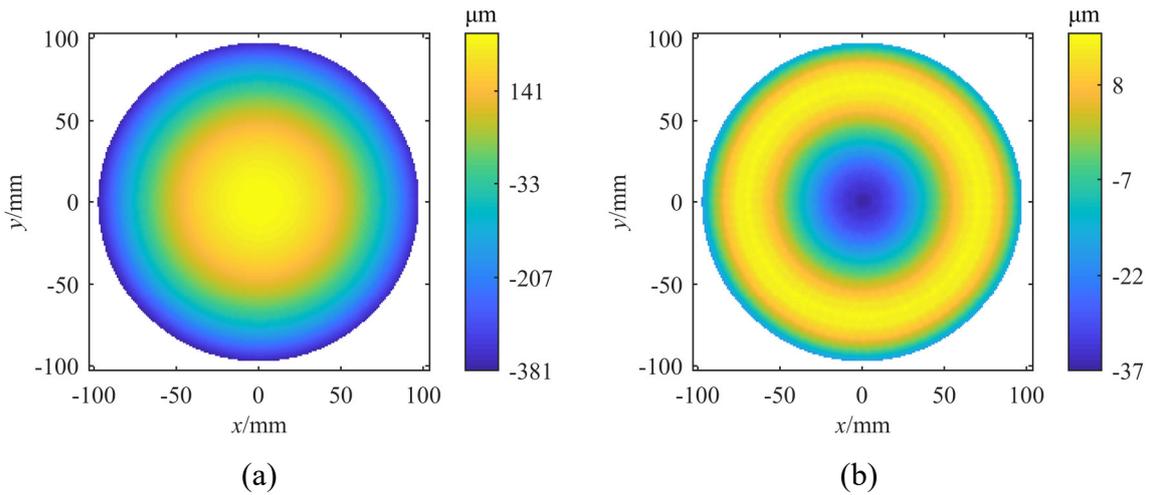

(a)                  (b)

Fig. 3 Film/substrate deformation (a) under residual stress of 800 MPa and (b) residual deviation from the best sphere

When the film was applied to different residual stress values, corresponding curvatures could be acquired from the finite element models. For the same plate, as mentioned in the energy method section, the curvature-stress relationship curve is shown in Fig. 4. By using the finite element method, it is very convenient to obtain the fitted curvature when the residual stress is known. However, it is very difficult to predict residual stress when the curvature is known. A trial and error method may be needed to obtain the residual stress inversely. It is pretty time-consuming and the finite element model needs to be reestablished once the dimensions of the plate (e.g. thickness, diameter) change.

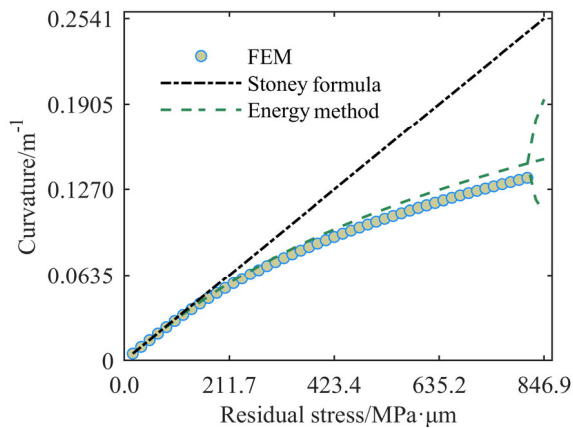

Fig. 4 Curvature-stress relationship using finite element method

**2.3 Modified Stoney formula**

The models to obtain the curvature-stress relationship using the energy method and finite element method are complex and time-consuming. Besides, the methods are designed to determine the curvature values when the residual stress is known. In actual scenarios, the curvature could be obtained by measurement and the residual stress in the film is desired to be determined from the curvature. The aforementioned models can't be applied directly and many curvature-stress points are needed to get the approximation value by interpolation. A formula like Stoney formula is desired so that the curvature and residual stress could be calculated easily from each other.

From the curvature-stress relationship curve in Fig. 4, we found that the exponential decay function with one coefficient was suitable to fit the curvature-stress points obtained by the finite element method. The function with one coefficient $p$ shown below was adopted. The residual stress $\sigma t$ is the independent variable. The residual stress $(\sigma t)_{Stoney}$ calculated by the Stoney formula is the dependent variable. Their relationship is shown in Fig. 5.

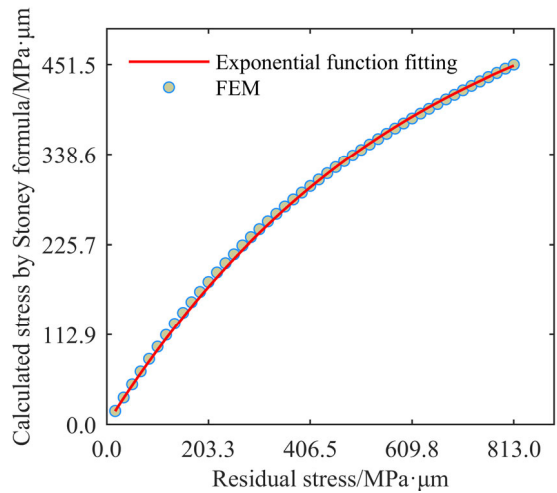

Fig. 5 Fitting of stress points with an exponential decay function

$$(\sigma t)_{Stoney}=(pe^{\frac{\sigma t}{p}}-p),\ (\sigma t)_{Stoney}=\frac{Eh^2}{6(1-v)}k \qquad (9)$$

The first-order derivative of the function with respect to $\sigma t$ could be easily calculated. When the $\sigma t$ is zero, the first-order derivative of the function is one, indicating that the calculated stress by the Stoney formula was equal to the actual applied residual stress. It means that the function is applicable in the linear deformation range. The coefficient $p$ is a negative number. With the increase of residual stress $\sigma t$, the first-order derivative of the function becomes smaller. A high correlation coefficient ($R^2$=0.9993) was achieved in the fitting. The difference between fitted values and actual values is very small, indicating the potential feasibility of the formula (9).

If the coefficient $p$ could be determined by the dimensions and material properties of the film/substrate system, the formula (9) itself could be applied to relate the curvature and the residual stress in the film without performing mathematical modeling or finite element simulation. In the next sections, the relationship between the coefficient $p$ and the dimensions and material properties of the film/substrate system was studied.

## 3. DETERMINATION OF COEFFICIENT

### 3.1 Relationship between coefficient *p* and substrate dimension

The coefficient *p* is introduced to describe the curvature-stress relationship when the film/substrate system is in large deformation. It is a reasonable prediction that the coefficient *p* is linked to the significance of large deformation. The significance of large deformation could be evaluated by the following parameter which is proposed by Finot et al.[9]

$$A = \frac{D^2}{h^3} \sigma t \quad (10)$$

For a specific film/substrate system with fixed dimension sizes, a value of the coefficient *p* could be obtained by fitting when the value of the residual stress $\sigma t$ changes. To find the relationship between the coefficient *p* and the plate dimension, four film/substrate systems with different thicknesses and diameters were selected to get the *p* values. Young's modulus of the system is set to be 100 GPa and the Poisson's ratio is 0.2. Their relationship is shown in Fig. 6.

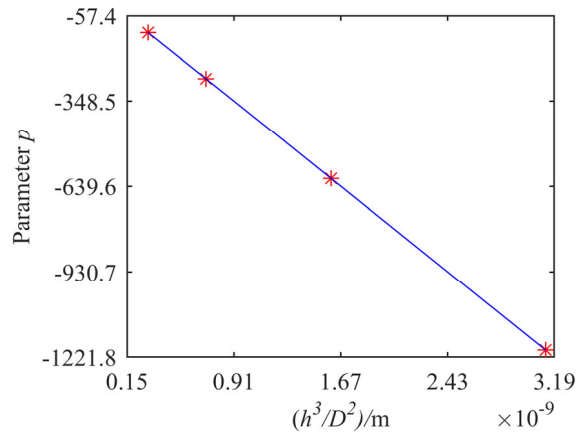

Fig. 6 Relationship between coefficient *p* and plate dimension

It is obvious that the coefficient *p* is linearly correlated to the $h^3/D^2$. A linear fitting was performed and the correlation coefficient $R^2$ was over 0.9999. Their relationship could be represented by a linear equation.

$$p = -\frac{h^3}{D^2} a + b \quad (11)$$

The slope of the linear function is $-a$ and the intercept value is *b*. The value of *b* is very small.

Therefore, the equation was simplified by discarding b.

$$p = -\frac{h^3}{D^2}a \quad (12)$$

The linear function was used for fitting and the correlation coefficient $R^2$ was still over 0.999, indicating that the coefficient p was negatively proportional to the $h^3/D^2$.

## 3.2 Relationship between coefficient p and substrate Young's modulus

Young's modulus is one of the main material property parameters to determine the film/substrate deformation under specific residual stress. Different Young's modulus values were chosen to find the relationship between the coefficient p and Young's modulus while other parameters (Poisson's ratio, thickness, and diameter of substrate) were fixed. Their relationship is shown in Fig. 7. The Poisson's ratio of the system was set to 0.2. The diameter of the system was set to 0.2 m and the thickness was set to 400 μm.

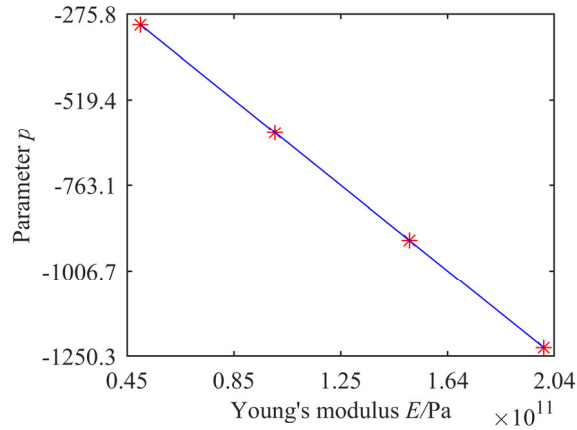

Fig. 7 Relationship between coefficient p and Young's modulus

It is apparent that there exists a strictly linear relationship between the coefficient p and Young's modulus. Linear fitting was conducted and the correlation coefficient $R^2$ was over 0.9999. Their relationship could be represented by a linear equation.

$$p = -\frac{h^3}{D^2}Ec + d \quad (13)$$

The slope of the linear function is $-h^3c/D^2$ and the intercept value is d. The value of d is very small. Therefore, the equation was simplified by discarding d.

$$p = -\frac{h^3}{D^2}Ec \quad (14)$$

The linear function (16) was used for fitting and the correlation coefficient $R^2$ was still over 0.9999, indicating that the coefficient $p$ was negatively proportional to Young's modulus $E$.

**3.3 Relationship between coefficient p and substrate Poisson's ratio**

For a specific film/substrate system, the last parameter is the Poisson's ratio of the substrate to determine the deformation under residual stress. Much more Poisson's ratio $v$ values which encompassed a wider range were selected to find their relationship. The relationship between Poisson's ratio $v$ and the coefficient $p$ was plotted in Fig. 8. The Young's modulus of the system was set to 100 GPa. The diameter of the system was set to 0.2 m and the thickness was set to 400 μm.

The linear correlation between the two variables was not significant. Then the relationships between the coefficient $p$ and $1/(1-v)$, $1/(1-v^2)$ were analyzed and plotted in Fig. 9 and Fig. 10. Although there exist strong linear correlations between the coefficient $p$ and $1/(1-v)$, $1/(1-v^2)$ the correlation coefficient in Fig. 10 was higher than that in Fig. 9. The coefficient $p$ was assumed to be linearly linked with $1/(1-v^2)$.

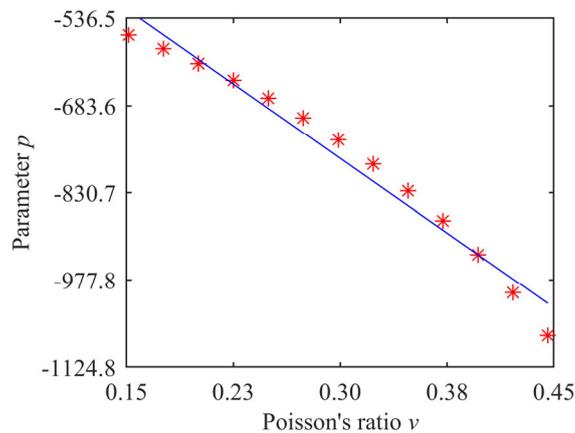

Fig. 8 Relationship between coefficient $p$ and Poisson's ratio $v$

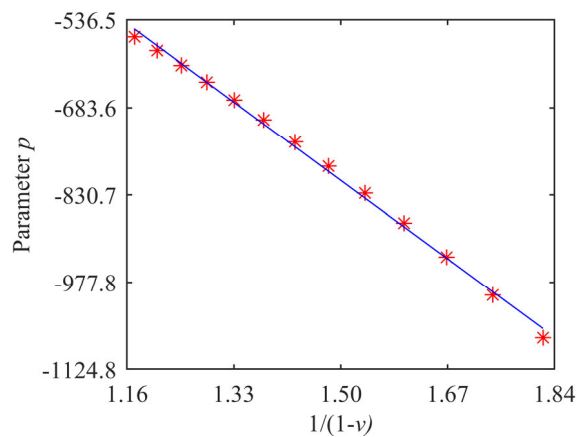

Fig. 9 Relationship between coefficient $p$ and $1/(1-v)$

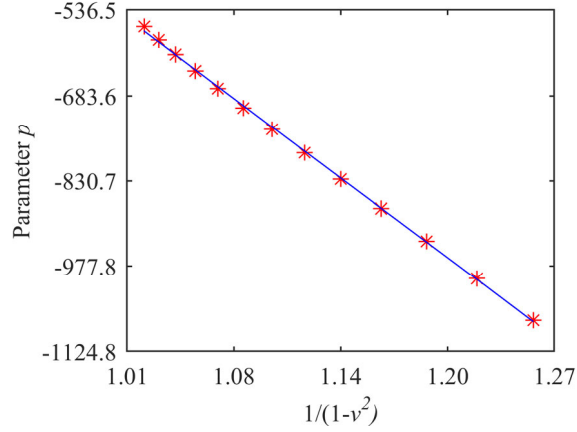

Fig. 10 Relationship between coefficient $p$ and $1/(1-v^2)$

Linear fitting was conducted and the correlation coefficient $R^2$ was over 0.999. Their relationship could be represented by a linear equation.

$$p = -\frac{h^3}{D^2} E\left(\frac{1}{1-v^2} m + n\right) \quad (15)$$

The coefficients $m$ and $n$ were obtained as 13.5 and 10.3, respectively. Then the parameter could be determined as:

$$p = -\frac{h^3}{D^2} E\left(\frac{13.5}{1-v^2} - 10.3\right) \quad (16)$$

Therefore, the modified Stoney formula could be expressed as follows by combining Equation (9):

$$\sigma t = p \ln\left(\frac{kEh^2}{6p(1-v)} + 1\right), \quad p = -\frac{h^3}{D^2} E\left(\frac{13.5}{1-v^2} - 10.3\right) \quad (17)$$

For a given film/substrate system, once the dimensions ($D$, $h$) and material properties ($E$, $v$) are known, the residual stress could be easily calculated using the above formula. The calculation only involves basic mathematical operations and no knowledge of advanced mathematics is needed. In the actual industrial applications, the engineers just need an ordinary scientific calculator to obtain the residual stress in the film from the curvature information without needing to employing plate deformation theories. Besides, the formula could be easily integrated into the software in which the residual stress could be calculated directly from measured substrate deformations.

### 3.4 Verification of modified Stoney formula

In the process of determining the relationship between the coefficient $p$ and the dimensions and the material properties of the film/substrate system, the values were selected artificially for the best presentation of their relationship. In this section, two actual film/substrate systems with true material property values were selected to verify the modified Stoney formula. Quartz glass was adopted as the first material. Young's modulus of Quartz glass is 72 GPa and Poisson's ratio is 0.17. The thickness of the substrate was 350 μm and the diameter was 150 mm. A comparison of different curvature-stress curves obtained by different methods is shown in Fig. 11.

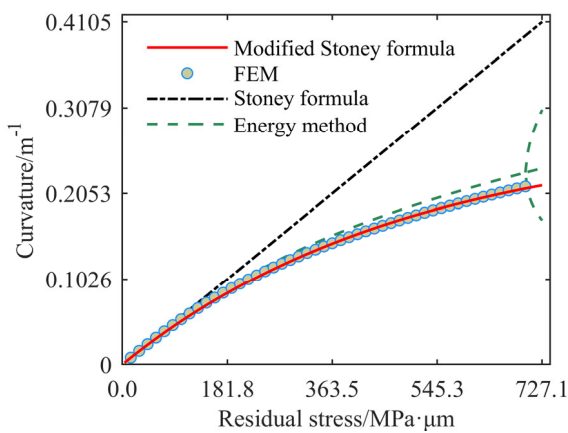

Fig. 11 Comparison of different curvature-stress curves obtained by different methods (Quartz glass)

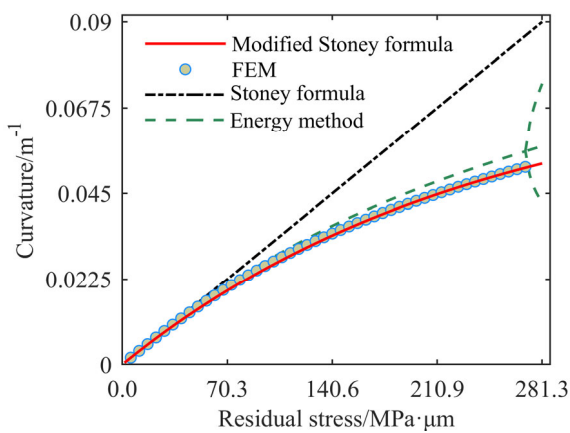

Fig. 12 Comparison of different curvature-stress curves obtained by different methods (Steel)

Steel was adopted as the other material. Young's modulus of steel is 210 GPa and Poisson's ratio is 0.3. The thickness of the substrate was 250 μm and the diameter was 250 mm. A comparison of different curvature-stress curves obtained by different methods is shown in Fig. 12. As shown in the figures, the error of the Stoney formula increases evidently with the increase of residual stress. The curvatures obtained by the energy method and finite element method are similar under the same residual stress. The modified Stoney formula successfully presents the curvature-stress relationship which is nearly identical to that obtained by the finite element method.

The study of the relationship between the coefficient $p$ and the dimension and material property of the film/substrate system was also conducted based on the curvature-stress data points based on the energy method. The same form as formula (15) was obtained. Only the coefficients $m$ and $n$ were different from those in formula (16). Since there are fewer assumptions in the finite element method than the energy method, the formula (17) is recommended for use. If there are more accurate models available, the modified Stoney formula could be updated using the same approach proposed in this paper.

## 4. CONCLUSIONS

Significant errors would be introduced for large deformations of film/substrate systems because of the assumption of small deformation in the Stoney formula. In this study, a modified Stoney formula was proposed to extend its scope of use to the nonlinear domain. One-phase exponential decay function with one coefficient $p$ is found to be suitable to relate the curvature of the substrate to the stress within the film. The coefficient $p$ of the modified Stoney formula was obtained by fitting with different values of the dimension and material property parameters of the film/substrate system. The coefficient $p$ is proportional to Young's modulus $E$ and cube of thickness $h^3$. It is inversely proportional to the square of the diameter $D^2$. The coefficient $p$ was linearly linked with the $1/(1-v^2)$.

The modified Stoney formula could be used to calculate the residual stress $\sigma t$ in the film directly from the measured curvature of the film/substrate system. The calculation involves only basic mathematical operations and could be easily performed using an ordinary scientific calculator. The modified Stoney formula could be applied for isotropic circular plates with various dimensions and materials.

## ACKNOWLEDGMENTS

The authors would like to acknowledge the financial support from the National Natural Science Foundation of China (51805135).

## DATA AVAILABILITY

The data that support the findings of this study are available from the corresponding author upon reasonable request.